\documentclass[10pt,amsmath,amssymb,latexsym]{article}
\usepackage{graphicx}  
\usepackage{dcolumn}   
\usepackage{bm}        
\usepackage{verbatim}  

\newtheorem{myDefinition}{Definition}
\newtheorem{myPostulate}{Postulate}

\newtheorem{myExample}{Example}

\begin{document}

\title{The general invariance of physical laws}
\author{Hitoshi Inamori\\
\small\it Soci\'et\'e G\'en\'erale\\
\small\it Boulevard Franck Kupka, 92800 Puteaux, France\\
}

\bigskip

\date{\today}

\maketitle

\begin{abstract}
Physical laws are a set of rules in the relationship between observations made by the experimenter. All these observations are made through a mechanism that links the external world to the experimenter's awareness, a mechanism which is not under the experimenter's control. We discuss how this mechanism can affect the physical laws as implied by the experimenter, and define fundamental physical laws as the ones which remain invariant under the change of such mechanism.
 
\bigskip


\end{abstract}

\section{Introduction}  

Suppose that an experimenter checks the violation of the Bell Inequality~\cite{Bell} by doing an Aspect type of experiment using a pair of entangled photons~\cite{Aspect}. We assume that each photon is detected by a pair of detection units after going through a beam splitter whose angle is set randomly by the setup. If the experimental setup is trustworthy, the outputs produced by the setup should violate the Bell Inequality, demonstrating that laws of Nature cannot be based on local hidden variables.

Now, suppose that the experimental setup is -- unbeknownst to the experimenter -- tampered with by an adversary so that it produces results that do obey Bell Inequality. This is possible, for instance, simply by having the experimental setup produce a predetermined flow of results that do obey Bell Inequality, independently of what the photons actually do. Indeed, the experimenter does not see the photons directly himself.

For an experimenter victim of such forgery, the laws of quantum mechanics will appear as flawed, and physical theories based on local hidden variables will remain a possibility.

Assuming that the above adversary is powerful enough to tamper with all the experimental results within what is allowed by the true physical laws, is there any way for the experimenter to infer the correct physical laws from what he observes?

The question can be turned around: Is it possible that the genuine laws of physics are based on local hidden variables, but that an adversary is tampering with all our experimental results to make us believe that the physical laws, as we perceive them, rule out local hidden variables?

The above is an illustration of the question asked in this paper: We construct our understanding of the physical laws based on our observations of the world: From the observed events, we imply relationships, or patterns, between these events, and we believe that some of these patterns, or rules, are universal. Now the observations are done through a certain mechanism, such as the experimental setup above, that translates the inputs from the external world, into observations witnessed by our awareness. To what extent can this translation mechanism affect the physical laws implied from our observations? What features of the physical laws remain invariant if the above translation mechanism is modified? 

The purpose of this paper is to formulate the above question in a more precise and general framework. To start with, we need to clarify what we mean by``actual observations'' witnessed by an experimenter. The discussion above forces us to remain as open minded as possible about the mechanism leading to the experimenter's observations, and our definition for the experimenter's observations will reflect this. We propose a setup that could explain how these observations are made, and within this framework, we shall formulate the question above more precisely, and discuss about the invariance of fundamental physical laws under change of the above translation mechanism.

\section{The actual observation witnessed by an experimenter}

In an experiment involving photons as the experiment discussed above, we usually refer to ``detection of a photon by a photo detection unit'' as experimental observations. We take for granted that when the experimental setup reports that such or such photo detection unit has triggered, that means -- or at least that indicates a plausibility -- that an entity called “photon” has hit such or such detection unit.

However, we have pointed out in the introduction that an adversary may have tampered with the experimental unit so that the experimenter is given completely bogus data for the experiment's results. For instance, regardless of the behavior of the photons and the actual triggering of the photo detectors, the computer running the experimental setup may have been programmed maliciously to give altered results. Besides, we cannot even assume that the photo emitter and photo detectors are working correctly, as any checking device, in the end, could be controlled by a sufficiently powerful adversary. There might not even be real photons in the experiment after all!

Being rigorous implies a certain level of paranoia as discussed above. In the above experiment, the computer screen readout is the only observation that the experimenter can make without ambiguity. All the implications about the so-called photons, photo detection units or beam splitters are interpretations based on some presumption. They are not actual observations.
Taking this into account, we define the actual observation made by an experimenter as follows:

\begin{myDefinition}[Actual observation made by an experimenter]  The {\bf actual observation made by an experimenter} is the set of all the observations or events that are directly and unambiguously witnessed by the experimenter. They exclude any inference or implication made from the observations.
\end{myDefinition}

In the above example, the readout from the computer screen, i.e. the visual signal coming from the screen that is directly available to the experimenter's awareness, is part of the actual observation made by the experimenter. The emission and detection of photons are not part of the actual observations made by the experimenter. They are implied events, based on the assumption that the experimental setup works as presupposed by the experimenter.

Now, we define the actual observation made by an experimenter as the set of all the data perceived by the experimenter. Suppose that the experimenter checks the result of a previous experiment in his notebook. The reading of a notebook is itself a physical experiment, and its outcome is part of the actual observation made by the experimenter. Previous experiment results are not guaranteed truth. They are stored in a book, a digital storage, a memory, and another physical process must occur to retrieve this information. There is nothing that guarantees that a sufficiently powerful adversary may not have tampered with this whole book keeping process. As such, the memory retrieval witnessed by an experimenter must be considered as part of the actual observation made by the experimenter. Past experimental results are not an intrinsic truth and is dependent on a physical process, like any actual observation made by the experimenter.

\section{The setup leading to the actual observation made by an experimenter}

Now that we have defined what the actual observations made by an experimenter are, we are ready to propose a setup which could lead to these observations. We assume in this paper that the actual observations are made in the setup described below:

\begin{myPostulate}[the Setup Postulate] There exists an {\bf external world}, an object which is not the experimenter himself, and which is being observed by the experimenter. The external world follows a certain set of laws, which we call the {\bf genuine physical laws}. All the actual observations are made by the experimenter through a mechanism, which is part of the external world, which links the external world to the experimenter's observations. This mechanism, called {\bf translation mechanism}, follows itself the genuine physical laws.
\end{myPostulate}

Note that everything that is perceived by the experimenter goes through the translation mechanism. The experimenter does not know the genuine physical laws, and ignores how the translation mechanism works. The only promise is that the translation mechanism obeys the genuine physical laws.

This postulate makes explicit what we usually accept implicitly without further questioning: the existence of a world outside the experimenter's awareness (i.e. we choose to rule out explanations such as ``everything is in your head''), a set of rules that applies universally in this world, and a causal link between the external world's phenomena and the observations made by the experimenter.

Note that the above postulate may not describe the truth: The external world might not exist, and even if it exists, it might not be endowed with any set of physical laws. However, as far as the experimenter is concerned, the above setup represents a possible explanation on how the actual observations are made. For now, we will assume that the Setup Postulate is the correct description of how experimental observations are made.

As the discussion above is rather abstract, let's consider a few specific and concrete examples: 

\begin{myExample} Consider the first example discussed in the introduction: in this situation, the genuine physical laws rule out local hidden variables, but the translation mechanism, i.e. the tampered experimental setup, yields actual observations that make the experimenter believe that the physical laws are compatible with local hidden variables.
\end{myExample}

Recall that the translation mechanism is only limited by the genuine physical laws. The observer does not know a priori what the genuine physical laws are, and has, at least, no direct way to check what is or is not allowed by the genuine physical laws. Every observation that the experimenter makes must go through the translation mechanism, and there is no way for the observer to examine this mechanism directly, i.e. bypassing the translation mechanism.

The following example is a straightforward illustration of this point:

\begin{myExample} Assume that all the inputs to the experimenter are restricted to a textual dialogue with a given interlocutor. The experimenter has no other access to the external world. We assume that this interlocutor has a computational power that is only limited by the laws of physics. Is the interlocutor able to make the experimenter believe in physical laws that are different from the genuine physical laws?  To what degree of liberty can the interlocutor deceive the experimenter?
\end{myExample}

In practice, the situations described in the examples above have never been implemented, and one may wonder whether they are theoretically realizable at all (how can we restrict a person to get inputs only through verbal communication?).
However, we also have an example which is much more familiar and which we do experience in practice quite often. 

\begin{myExample} During a dream, the inputs to the experimenter awareness, and the resulting implied laws of physics, are altered compared to what the experimenter witnesses when he is awake. The actual observation made by the observer, however, remains a result of a mechanism (brain activity) that does follow the actual laws of physics, even during a dream. Moreover, some reader will also agree that the content of the memory can be different during a dream, which illustrates the idea that past events may not be an immutable truth, but depend on the mechanism that links the external world to the experimenter's actual observation.
\end{myExample}

\section{The general invariance of physical laws}

We are now ready to come back to the main question of this paper.

The experimenter does make actual observations. We assume that the Setup Postulate describes correctly how these observations are made. However the experimenter does not know the genuine laws of physics and does not know what is the translation mechanism which led to his observations. The translation mechanism is only limited by the genuine laws, and depending on the translation mechanism, the experimenter could have been led to believe in different implied physical laws. This is what is illustrated in the example of the introduction.

Alternatively, let's suppose that there is another experimenter. The experimenters are unable to check whether they see the external world through the same translation mechanism. If the two experimenters make their respective observations through distinct translation mechanisms, they may end up with implied physical laws that are also distinct.  

Nevertheless, could one find common denominators between the physical laws implied by different translation mechanisms?

It would be tempting or somewhat natural to impose some condition on the translation mechanisms allowed in the question above. For instance, it would seem natural to assume that such translation mechanism be ``bijective'', or at least, that no ``information'' is lost (or created) through the change of the translation mechanism. However, defining such constraint is not trivial (how do we define “information”?) nor does it seem necessary. The only constraint is that the translation mechanism is allowed by the external world and the genuine physical laws. One may argue that a translation mechanism that cuts out nearly all the information from the external world would trivially lead to an implied physical law that is empty and with little relevance (note that we assumed that the experimenter does observe something, so we cannot cut absolutely all the information). But is such translation mechanism truly feasible? Is it possible to impoverish in practice the quantity of information that an experimenter gets? We could throw the experimenter in a prison cell without any communication with the outer world. Does this mean that the experimenter does not experience anything inside the cell? Remember that even in a dream, where the experimenter is cut out from a big part of the data from the outer world, the experimenter continues witnessing events, in a relatively detailed manner.

We do not have any direct answer as to what the common denominators shared by the implied physical laws are. Rather, we state that the fundamental features of physical laws are the ones that do not depend on the choice of the translation mechanism. 

\begin{myDefinition}[Fundamental features of physical laws] Assuming that the Setup Postulate is correct, the fundamental features of physical laws are the features of the implied physical laws that are invariant under the change of the translation mechanism.
\end{myDefinition}

Said simply, translation mechanisms provide different ways of looking at the same external world, animated by the same genuine physical laws. Our proposition states that the fundamental laws of nature should be invariant whatever this way of looking is.

As noted above, we do not have concrete answer as to what the fundamental features of physical laws are. However, we can try to find out what do not constitute fundamental features of physical laws. According to our definition, a law that is not invariant under the change of the translation mechanism is not a fundamental one. 

In the example of the introduction, the change of the translation mechanism consisted in fiddling with the experimenter's computer. Such tampering seems physically feasible. Does it mean that violation of the Bell Inequality is not a fundamental feature of the physical laws? It may be so, if all the experiments that do test Bell Inequality in one way or another can all be tampered with by the adversary. But is such general tampering really possible?

A more straightforward situation was the hypothetical situation in which all the inputs to the experimenter are restricted to a textual dialogue with a given interlocutor, who has unlimited computational resources. In such a case, we are tempted to think that the experimenter can be led to believe in any implied physical law, and pretty much any feature of physical laws (and in particular, the equations describing physical laws and the physical constants) could be considered as being not fundamental. But is it really possible to restrict all communication with the experimenter to textual dialogue? How could the experimenter learn the meaning of the text, if he had no access to other form of information?

\section{Discussion}

The idea that the inputs to an experimenter brain may be completely fabricated by a powerful adversary is certainly not new. Descartes, among others, had suggested the existence of an omnipotent ``Evil Demon'' who can mislead the observer by presenting a complete illusion of the external world~\cite{Descartes}. The goal of this paper is not to ask whether such Evil Demon exists, but to ask what can be said about the fundamental laws of nature, given that we cannot exclude the existence of such Demon.

Our proposal is that if an external world exists, that is endowed by universal physical laws, and that any experimental observation is the product of a translation mechanism that is allowed by these universal laws, then the physical laws implied by the experimenter depends on the translation mechanism. Our point is that if two different translation mechanisms lead to two different implied laws of physics, then such implied laws are not the fundamental ones. 

The question is: Are the current laws of physics, with their mathematical representations and constants, fundamental according to our definition? 

If they are, it means that either our Setup Postulate is wrong, or that the translation mechanisms are restricted to the ones that lead to the implied laws of physics as we know them. In the latter case, what causes this restriction?

Alternatively, are our common laws of physics not truly fundamental according to our definition? In this case, how do we explain that many different experimenters of the world seem to agree on a unique set of physical laws?  

One possible explanation would be that the shared physical laws correspond to common features shared by these experimenters' translation mechanisms. An experimenter subject to a completely distinct translation mechanism is likely to infer physical laws that are very different from ours.


\end{document}